\documentclass[aps,prl,reprint,amsmath,nofootinbib,preprintnumbers]{revtex4-2}

\usepackage{graphicx}
\usepackage{bm}
\usepackage{amssymb}

\usepackage{tikz}
\newcommand{\Plus}{{\mathord{\tikz\draw[line width=0.2ex, x=1ex, y=1ex] (0.5,0) -- (0.5,1)(0,0.5) -- (1,0.5);}}}

\usepackage{pifont}

\def\Nf{N_{\rm f}}

\def\vareps{\varepsilon}
\def\qhat{\hat q}

\def\min{{\rm min}}

\def\BH{{\rm BH}}

\def\LPM{{\rm LPM}}
\def\LPMplus{{{\rm LPM}\Plus}}
\def\Elpma{E_{\rm LPM}}
\def\Elpm{\hat E_{\rm LPM}}
\def\kgamma{k_\gamma}
\def\tform{t_{\rm form}}
\def\Sxtra{S_{\rm extra}}
\def\pair{{\rm pair}}
\def\me{m}
\def\mgamma{m_\gamma}
\def\Re{\operatorname{Re}}

\def\Mo{{\bar M}_0}

\def\gammaE{\gamma_{\rm\scriptscriptstyle E}}
\def\yfrak{\eta}
\def\lnGamma{\operatorname{ln\Gamma}}

\def\PSI{\bar\psi}

\begin{document}

\title{Extremely high-energy bremsstrahlung in matter}

\author{Peter Arnold}
\author{Joshua Bautista}
\affiliation{%
    Department of Physics,
    University of Virginia,
    Charlottesville, Virginia 22904-4714, USA
}%
\author{Omar Elgedawy}
\affiliation{%
  State Key Laboratory of Nuclear Physics and
  Technology, Institute of Quantum Matter, South China Normal
  University, Guangzhou 510006, China
}
\affiliation{%
  Guangdong Basic Research Center of Excellence for
  Structure and Fundamental Interactions of Matter, Guangdong
  Provincial Key Laboratory of Nuclear Science, Guangzhou
  510006, China
}
\affiliation{%
  CPHT, CNRS, \'{E}cole polytechnique, Institut Polytechnique de Paris,
  91120 Palaiseau, France
}
\author{Shahin Iqbal}
\affiliation{%
National Centre for Physics,
  Shahdra Valley Road,
  Islamabad, 45320 Pakistan
}%
\affiliation{%
  Theoretical Physics Department,
  CERN,
  CH-1211 Geneva 23, Switzerland
}%

\preprint{CPHT-RR013.042026}
\preprint{CERN-TH-2026-091}

\date {\today}

\begin{abstract}%
The theory of bremsstrahlung $e \to e\gamma$
by extremely high energy electrons passing through ordinary matter
has been qualitatively incomplete.
We revisit the suppression of bremsstrahlung by the
Landau-Pomeranchuk-Migdal (LPM) effect, here accounting for
quantum disruption of that effect from pair production.
Our analysis covers the full range of ultra-relativistic electron
and photon energies (subject to a few simplifying approximations).
\end{abstract}

\maketitle



\begin {figure}
  \includegraphics[scale=0.55]{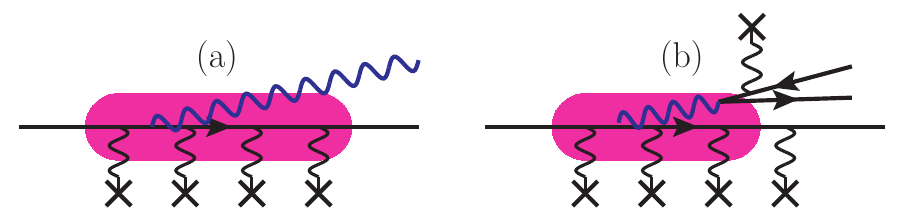}
\caption{
     \label{fig:overlap}
     (a) Multiple scattering during the bremsstrahlung formation time,
     depicted by the length of the shaded region.
     (b) The bremsstrahlung formation time disrupted by
     pair production $\gamma \to e\bar e$.
  }
\end {figure}

Very high energy electrons or positrons passing through ordinary matter
lose energy predominantly through showering via repeated
bremsstrahlung ($e \to e\gamma$) and pair production ($\gamma \to e\bar e$)
mediated by the electric fields inside atoms.
The bremsstrahlung rate for a high-energy electron passing through the
Coulomb field of a single atomic nucleus was
first calculated by Bethe and Heitler in 1934 \cite{BH}.
In 1953, Landau and Pomeranchuk noted that the quantum duration of
bremsstrahlung (known as the formation or coherence time $\tform$)
grows with
electron energy and eventually becomes larger than the mean free time
between elastic scatterings from the medium \cite{LP1,LP2,LPenglish}.
They found that multiple elastic scattering during the formation
time, as schematically depicted in fig.\ \ref{fig:overlap}a,
can cause large suppression of the net bremsstrahlung rate at
high energy, depending on the initial electron energy $E$ and bremsstrahlung
photon energy $\kgamma$.  A few years later, Migdal \cite{Migdal} made
the first fully quantum-mechanical calculation of this suppression,
now known as the Landau-Pomeranchuk-Migdal (LPM) effect.
The LPM effect has been
well studied experimentally \cite{SLAC1,SLAC2,CERNlpm1,CERNlpm2}.
(See \cite{SpencerReview} for a comprehensive pre-2000 review.)
Parametrically, the LPM bremsstrahlung rate is suppressed compared to the
Bethe-Heitler rate as
\begin {equation}
   \left[ \frac{d\Gamma}{d\kgamma} \right]_\LPM
   \sim
   \min\left( 1 \,,\, 
      \sqrt{ \frac{\kgamma \Elpma}{E(E-\kgamma)} }
   \,\,\right)
   \times
   \left[ \frac{d\Gamma}{d\kgamma} \right]_\BH ,
\label {eq:LPMrate0}
\end {equation}
where $\Elpma$ is a medium-dependent energy scale that is, for example,
$\Elpma \simeq 2.5$ TeV for Gold and $234$ PeV for air \cite{SpencerReview}.
Accelerator experiments \cite{SLAC1,SLAC2,CERNlpm1,CERNlpm2} have
been limited to electron energies much smaller than $\Elpma$ and
so have studied
the LPM effect for $\kgamma \ll E \ll \Elpma$. 

In 1964, Galitsky and Gurevich \cite{Galitsky} noted that at
extremely high energy there are regions of $(k_\gamma,E)$ where
the formation time for
bremsstrahlung is large enough to encompass not only multiple scattering
from the medium but \textit{also} the
subsequent disappearance of the bremsstrahlung photon via medium-induced
pair production, as depicted in fig.\ \ref{fig:overlap}b.
They argued qualitatively that premature disappearance of the
photon would cause even further suppression of (\ref{eq:LPMrate0})
by an additional factor of
$\Sxtra \sim \min\bigl(1,(\Gamma_\pair\tform)^{-1}\bigr)$,
where $\Gamma_\pair$ is the pair production rate and
$\tform$ is the bremsstrahlung formation time if the effects
of pair production on bremsstrahlung
are ignored (as in the original LPM calculations).
Recently \cite{softqed1}, revisiting this problem in the simplifying
limit $k_\gamma \gg \Elpma$ of extremely-high energy bremsstrahlung,
we argued that Galitsky and Gurevich were mistaken about the direction of the
effect.  We confirmed by a full analytic
calculation, in the spirit of Migdal's, that the effect
of subsequent pair production is to disrupt LPM suppression and so
significantly \textit{increase} the LPM rate compared to (\ref{eq:LPMrate0})
by an additional factor of order
\begin {equation}
   \Sxtra
   \sim
   \max\left( 1 \,,\, \Gamma_\pair\tform \right) ,
\label {eq:Sextra}
\end {equation}
which is large when $\tform \gg 1/\Gamma_\pair$.

Our previous analysis applies only to electron energies
larger than might be reasonably foreseen for future experimental tests
of (\ref{eq:Sextra}), and so we have now extended that analysis
to lower bremsstrahlung energies $\kgamma \lesssim \Elpma$.
This extension
requires including effects of the electron mass $\me$, which
we ignored in our earlier analysis.
For sufficiently small (but still very high
energy) $\kgamma$, one must also include the medium-induced mass $\mgamma$
of the photon, known in this context as the ``dielectric effect.''
This paper presents our full result and also discusses
simpler limiting formulas in various regions of $(\kgamma,E)$.
Details of the calculation are left to a
companion paper \cite{softqed3}.


\vspace{-0.1in}
\subsection*{Assumptions and Conventions}
\vspace{-0.1in}

We work in natural units $\hbar = c = 1$ and so will use
the terms formation time and formation length interchangeably.
For the sake of these first calculations of the effect of pair production
on bremsstrahlung (extended to cover the region $k_\gamma \lesssim \Elpma$),
we retain a number of simplifying approximations made in ref.\ \cite{softqed1}.
First, we assume that the medium is large compared to (and approximately
homogeneous over) the bremsstrahlung formation length.
We will not include edge effects from a high-energy electron
entering or leaving a finite medium.  Migdal was able to find analytic
results for the LPM effect by working to leading-log order
in a large logarithm $\ln(a_Z \me)$ that appears in the Bethe-Heitler
rate, where $a_Z$ is the Thomas-Fermi atomic radius
$a_Z \sim Z^{-1/3} a_{\rm Bohr} \sim Z^{-1/3}/\alpha\me$ for atoms with
atomic number $Z$.  We will do the same regarding that particular
logarithm.  We also focus on the case $Z \gg 1$ and so ignore
all effects suppressed by $1/Z$.

The literature on the QED LPM effect typically expresses results in
terms of the radiation length $X_0$ of the medium --- the
distance over which a high energy particle (but not so high energy
that the LPM effect is important) loses all but $1/e$ of its energy.
In the simple case of a single-element material of non-light atoms,
$X_0 \simeq [4nZ^2\alpha^3 \ln(184 Z^{-1/3}) / \me^2]^{-1}$, where $n$ is the
density of atoms and the
logarithm is an improved version of $\ln(a_Z\me)$
adjusted to be somewhat better than merely a leading-log
approximation \cite{SpencerReview}.
The energy scale $\Elpma$ introduced earlier is given by
$\Elpma = \alpha\me^2 X_0/4\pi$ (using the normalization convention
of \cite{SpencerReview,RPP2024}).
The techniques that we use for our calculation \cite{softqed1,softqed3}
are adapted from the literature on the QCD LPM effect, where it is
more convenient to characterize the medium in terms of a parameter
$\qhat$ defined as the proportionality constant
$(\Delta p_\perp)^2_{\rm typical} \simeq \qhat \,\Delta t$
for a random walk of the total transverse momentum kick
$\Delta p_\perp$ that a high-energy particle picks up in a time $\Delta t$
traversing the medium.
In applications to the LPM effect, there
is a very mild logarithmic dependence of $\qhat$ on
$(\kgamma,E)$ which, for the energy ranges explored here,
changes $\qhat$ by at most a factor of 2.  We find it
convenient to express our results in terms of $\Elpm \equiv \me^4/\qhat$.
At lower energies, our $\Elpm$ is related to the conventional $\Elpma$
by $\Elpm = 2\Elpma$.  (See ref.\ \cite{softqed1} for a more thorough
discussion.)
For simplicity of calculation and presentation, we will not keep track
of the mild energy dependence of $\qhat$
(following ref.\ \cite{softqed1}).
Other than this approximation, we
keep track of all other logarithmic dependencies in our calculations.

Only purely QED processes are included in our analysis.
This is not an issue for photon energies below roughly $10^{20}$ eV
\cite{LPMphotoKlein},
beyond which photo-nuclear processes
($\gamma A \to$ hadrons) can become more important than
pair production $\gamma\to e\bar e$ for
disrupting LPM suppression of bremsstrahlung
\cite{softqed3}.
One might also wonder about purely-QED
muon pair production $\gamma \to \mu\bar\mu$.
Though that could also be treated
with our methods, it is negligible compared to electron pair production
for photon energies below roughly
$(m_\mu/\me)^4 \Elpma \sim 10^9 \Elpma \gtrsim 10^{21}$ eV.

In our case of interest,
where subsequent pair production $\gamma\to e\bar e$
potentially occurs \textit{during} the bremsstrahlung process
$e\to e\gamma$ that produced the photon in the first place,
isolating a bremsstrahlung rate for $e\to e\gamma$ by
itself becomes more subtle.
We instead compute the
rate $d\Gamma/dx_\gamma$ for the initial electron to lose
a fraction $x_\gamma \equiv \kgamma/E$ of its energy $E$
through radiative processes
($e\to\gamma e$ or overlapping $e\to\gamma e\to ee\bar e$).
This rate has the advantage that it need not
draw any precise distinction between energy loss by bremsstrahlung
that (i) does or (ii) does not overlap subsequent pair production.
As discussed in ref.\ \cite{softqed1}, the combined rate is also easier to
calculate.

Throughout, our formulas for $d\Gamma/dx_\gamma$ will refer to
this differential rate for radiative energy loss.  However, in the case of
overlapping $e \to e\gamma \to ee\bar e$, there is a conceptual issue to
consider: with two electrons in the final state, which of them should
be used to determine the energy loss $E - E_e^{\rm final}$?
Overlapping $e \to e\gamma \to ee\bar e$
turns out \cite{softqed1,softqed3} to be
unlikely (suppressed by $\alpha$) unless $k_\gamma \ll E$, in which
case the pair-produced electron is soft and so can be unambiguously
distinguished
from the hard-electron daughter that is heir to the original electron.

In this paper, we refer to the initial high-energy particle
as an electron ($e \to e\gamma \to ee\bar e$), but the analysis
is the same for an initial high-energy positron
($\bar e \to \bar e\gamma \to \bar ee\bar e$).


\vspace{-0.1in}
\subsection*{Original LPM rate}
\vspace{-0.1in}

In what follows, we refer to the rate originally
computed by Migdal as the ``LPM'' rate and refer to our
calculation, which includes the possible effects of overlapping
pair production, as the ``$\LPMplus$'' rate.  Including the
dielectric effect, Migdal's result \cite{Migdal} was equivalent
\cite{softqed3} to
\begin {widetext}
\begin {equation}
  \left[ \frac{d\Gamma}{dx_\gamma} \right]_\LPM
  \simeq
  \frac{\alpha}{\pi}
  \Re\left\{
    i P_{e\to\gamma}(x_\gamma)
    \biggl[
      \Omega_0
      - \Bigl( \frac{\me^2}{2\Mo}{+}\frac{\mgamma^2}{2\kgamma} \Bigr)
        \,\PSI\Bigl(
          1 \,;
          \frac{\me^2}{4\Mo\Omega_0}{+}\frac{\mgamma^2}{4\kgamma\Omega_0}
        \Bigr)
    \biggr]
    + i x_\gamma \frac{\me^2}{2\Mo}
        \,\PSI\Bigl(
          \frac12 \,;
          \frac{\me^2}{4\Mo\Omega_0}{+}\frac{\mgamma^2}{4\kgamma\Omega_0}
        \Bigr)
  \right\}
\label {eq:LPMrate}
\end {equation}
\end {widetext}
where we define $\PSI(r;z) \equiv \psi(r{+}z) - \ln z$ and
$\psi(z) = \Gamma'(z)/\Gamma(z)$ is the digamma function.
Above, $\Mo \equiv (1{-}x_\gamma)E/x_\gamma$;
$\Omega_0 \equiv \sqrt{-ix_\gamma \qhat/2(1{-}x_\gamma)E}$;
the medium-induced photon mass
(dominated by forward Compton scattering from atomic electrons) is
$m_\gamma \simeq \sqrt{4\pi\alpha Z n/\me}$ ($m_\gamma\ll m$); and
$P_{e\to\gamma}$ is the unregulated
Dokshitzer-Gribov-Lipatov-Altarelli-Parisi (DGLAP) splitting function
$P_{e\to\gamma}(x_\gamma) = {\bigl[1 + (1{-}x_\gamma)^2\bigr]/x_\gamma}$.
Throughout this paper, one may use $\qhat = \me^4/\Elpm$ to
rewrite formulas involving $\qhat$ in terms of $\Elpm$.

\begin{figure*}
  \includegraphics[scale=1.0]{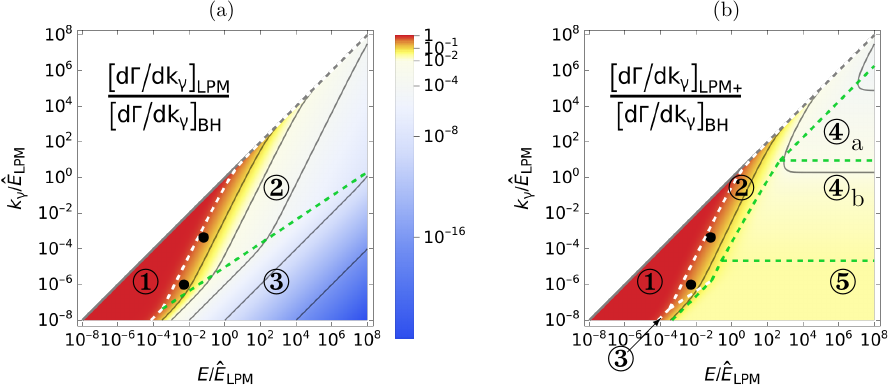}
  \caption{
     \label{fig:overBH}
     (a) Log-log-log contour plot of the ratio $\LPM/\BH$
     of the original
     LPM rate (\ref{eq:LPMrate})
     to the Bethe-Heitler rate (\ref{eq:BHrate}) vs.\ $E/\Elpm$ and
     $\kgamma/\Elpm$.  Note the non-uniform
     spacing $(10^{-1},10^{-2},10^{-4},10^{-8},10^{-16})$ of the contour lines.
     (b) The same, but now for $\protect\LPMplus/\BH$.
     Both plots use $\alpha = 1/137$ and $m_\gamma/\me = 10^{-4}$ (roughly
     the value for Lead or Gold).  Given the scaling of the axes, the plots are
     independent of the value of $\Elpm$.  The value of $m_\gamma/\me$ only
     affects the location of the dielectric-effect regions 3 and 5.
     To keep the plot as otherwise general as possible,
     we have not explicitly marked where
     the high energy approximations used
     [(a,b) $E \gg m$, (a) $\kgamma \gg m_\gamma$, and (b) $\kgamma \gg m$]
     break down.  (Nor have we
     indicated the low-energy region where the corresponding
     differential rate for
     energy loss from ionization dominates over that from radiation.
     See section IX of our companion paper \cite{softqed3}.)
     For reference, the two black dots indicate one illustrative
     example each of $(k_\gamma,E)$ from the range of values tested
     by two dedicated LPM experiments:
     {(5 MeV, 25 GeV)} in Gold \cite{SLAC1} and
     {(2 GeV, 287 GeV)} in Iridium \cite{CERNlpm1}, for which
     $\Elpm \simeq 2\Elpma \simeq 5.0$ and $4.5$ TeV respectively.
  }
\end{figure*}

For comparison, the Bethe-Heitler rate, which does not include
the LPM and dielectric effects, is approximately
\begin {equation}
  \left[ \frac{d\Gamma}{dx_\gamma} \right]_\BH
  \simeq
  \frac{\alpha \qhat}{6\pi\me^2} \,
  \bigl[ 2 P_{e\to\gamma}(x_\gamma) + x_\gamma \bigr] .
\label{eq:BHrate}
\end {equation}
Fig.\ \ref{fig:overBH}a
plots the ratio LPM/BH of the LPM rate
(\ref{eq:LPMrate}) to the Bethe-Heitler rate (\ref{eq:BHrate}).
This figure shows the suppression of the bremsstrahlung rate by medium effects
for $E \gg \Elpma$ or sufficiently small $\kgamma/E$. 
Migdal's formula (\ref{eq:LPMrate}) makes smooth transitions between the
three different regions of this graph, which are characterized in
the top half of table \ref{tab:regions} by the corresponding
limits of (\ref{eq:LPMrate})
deep inside each region.  The ``dielectric'' region represents
modification to the soft-photon limit of the Bethe-Heitler
rate (\ref{eq:BHrate}) by the medium-induced photon
mass $m_\gamma$.  The boundaries of these regions are summarized
parametrically in table \ref{tab:boundaries}.  In the
figure, we have adopted the more specific convention of drawing
boundary lines where the limiting
rate formulas (table \ref{tab:regions})
for two adjoining regions are equal to each other.

\begin {table}
\renewcommand{\arraystretch}{1.5}
\setlength{\tabcolsep}{3pt}
\begin {tabular}{cll}
\toprule
  \multicolumn{2}{l}{region} & \qquad $d\Gamma/dx_\gamma \simeq$
\\
\hline
  1. & BH &
    $  \frac{\alpha\qhat}{6\pi m^2}
       \bigl[ 2 P_{e\to\gamma}(x_\gamma) + x_\gamma \bigr]  $
\\
  2. & deep LPM &
    $  \frac{\alpha}{2\pi} \, P_{e\to\gamma}(x_\gamma)
       \sqrt{ \frac{x_\gamma\qhat}{(1{-}x_\gamma)E} }    $
\\
  3. & dielectric &
    $  \frac{2 \alpha \qhat x_\gamma}{3\pi m_\gamma^2}   $
\\[4pt]
\hline
\\[-11pt]
  4$_{\rm a}$. & deep $\LPMplus$ &
    $  \frac{\alpha}{2\pi} \, P_{e\to\gamma}(x_\gamma) \,
       \Gamma_\pair^{\rm(LPM)}
       \biggl[
         \ln\Bigl(
            \tfrac{2^{17/3}}{3\pi\alpha}
         \Bigr)
         + \tfrac16
      \biggr]                                         $
\\[6pt]
  4$_{\rm b}$. & deep $\LPMplus$ &
    $  \frac{\alpha}{2\pi} \, P_{e\to\gamma}(x_\gamma) \,
       \Gamma_\pair^{\rm(0)}
       \biggl[
         \ln\Bigl(
            \tfrac{m^2/\kgamma}{\Gamma_\pair^{\rm(0)}}
         \Bigr)
         + \tfrac{41}{21}
      \biggr]                                         $
\\[6pt]
  5. & dielectric$\Plus$ &
    $  \frac{\alpha}{2\pi} \, P_{e\to\gamma}(x_\gamma) \,
       \Gamma_\pair^{\rm(0)}
       \biggl[
         \ln\Bigl(
            \tfrac{m^2}{m_\gamma^2}
         \Bigr)
         + \tfrac{20}{21}
      \biggr]                                         $
\\[6pt]
\botrule
\end {tabular}
\caption{%
\label{tab:regions}%
  Limiting formulas (far from region boundaries) for the
  differential energy rate
  $[d\Gamma/dx_\gamma]_\LPM$ or $[d\Gamma/dx_\gamma]_{\protect\LPMplus}$
  for radiative energy loss
  in various regions of fig.\ \ref{fig:overBH}.
  Above, $\Gamma_\pair^{(\LPM)} \simeq (3\alpha/8)\sqrt{\qhat/\kgamma}$ is the LPM
  pair-production rate in the limit $k_\gamma \gg \Elpma$,
  and $\Gamma_\pair^{(0)} \simeq 7\alpha\qhat/18\pi m^2$ is the
  Bethe-Heitler pair production rate relevant to energies
  $\kgamma \ll \Elpma$.
}
\end {table}

\def\vbar{\,$|$\,}
\def\ph{\phantom{${}_{\rm b}$}}
\begin {table}
\renewcommand{\arraystretch}{1.5}
\setlength{\tabcolsep}{3pt}
\begin {tabular}{ll}
\toprule
  \ph 1\vbar2 &
    $ \kgamma \sim \frac{E^2}{\Elpma} $
    \quad (when $E \lesssim \Elpma$)
\\
  \ph 1\vbar3 &
    $ \kgamma \sim \frac{m_\gamma}{m} E $
\\
  \ph 2\vbar3 &
    $ \kgamma \sim
      \left(\frac{m_\gamma}{m}\right)^{4/3} \Elpma^{1/3} E^{2/3} $
\\[4pt]
\hline
\\[-11pt]
  \ph 2\vbar4$_{\rm a}$ &
    $ \kgamma \sim \alpha E \ln\bigl( \frac{1}{\alpha} \bigr) $
\\
  \ph 2\vbar4$_{\rm b}$ &
    $ \kgamma \sim \frac{(\alpha E)^2}{\Elpma}
       \ln^2\bigl( \frac{\Elpma}{\alpha^{3/2} E} \bigr) $
\\
  \ph 2\vbar5 &
    $ \kgamma \sim \frac{(\alpha E)^2}{\Elpma}
       \ln^2\bigl( \frac{m}{m_\gamma} \bigr) $
\\
  \ph 3\vbar5 &
    $ \kgamma \sim \alpha^{1/2} \frac{m_\gamma}{m}
       E \ln^{1/2}\bigl( \frac{m}{m_\gamma} \bigr) $
\\
  4$_{\rm b}$\vbar5 &
    $\kgamma \sim \left(\frac{m_\gamma}{m}\right)^2 \frac{\Elpma}{\alpha} $
\\[6pt]
\hline
  \multicolumn{2}{l}{
    Additional restrictions (high-energy limit):
  }
\\[-3pt]
  & $E \gg \me$ \hspace{1em} and \hspace{1em} $\kgamma \gg \me$
\\
\botrule
\end {tabular}
\caption{%
\label{tab:boundaries}%
  Parametric boundaries between the different regions of fig.\ \ref{fig:overBH}.
  For $E \gg \Elpma$, the 1\vbar2 boundary asymptotically approaches
  $k_\gamma \simeq E$ following $E-\kgamma \sim \Elpma$.
}
\end {table}

\vspace{-0.1in}
\subsection*{Inclusion of overlapping pair production}
\vspace{-0.1in}

\begin {figure}
  \includegraphics[scale=0.9]{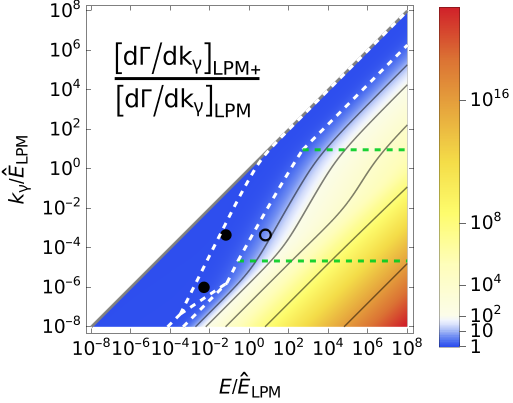}
\caption{
     \label{fig:LPM+overLPM}
     Log-log-log contour plot of the $\protect\LPMplus/\LPM$ ratio
     of the $\protect\LPMplus$ rate (which includes overlapping pair production)
     to the LPM rate (which does not) vs.\
     $E/\Elpm$ and $\kgamma/\Elpm$.
     The small open circle is explained in the conclusion.
  }
\end {figure}

In contrast, fig.\ \ref{fig:overBH}b
presents our results for the $\LPMplus/\BH$ ratio from
including the effects of overlapping pair production on the original
LPM bremsstrahlung rate.
The full formula for the $\LPMplus$ rate, analogous to the original LPM rate
(\ref{eq:LPMrate}), is given in the appendix.
Region 4$_{\rm a}$ is the region that was previously studied in
ref.\ \cite{softqed1}, where the electron mass $m$ and dielectric effect
could be ignored.  The full result presented here now extends that analysis
to regions 4$_{\rm b}$ and 5.  The relatively simple
limiting forms of our full $\LPMplus$ rate deep inside these regions,
and parametric formulas for the new boundaries between regions, are
included in tables \ref{tab:regions} and \ref{tab:boundaries}.
By comparing figs.\ \ref{fig:overBH}a and b, we see that pair production
indeed arrests the rapid fall of the LPM rate in these regions.
Fig.\ \ref{fig:LPM+overLPM} plots the $\LPMplus/\LPM$ ratio to directly
display the enhancement of the LPM rate.

At \textit{leading-log} order,
the behavior of the limiting rate formulas in Table
\ref{tab:regions} have a simple physical interpretation for regions 4
and 5.  Those formulas can be understood as the medium-induced
$\gamma{\to}e\bar e$ pair-production rate $\Gamma_\pair$
times an effective Weizs\"acker-Williams probability distribution
$(\alpha/2\pi)\,P_{e\to\gamma}(x_\gamma)\,\log(\cdots)$ for finding a photon
component ($e{\to}e\gamma$)
of the initial high-energy electron.
We did not build this picture into our calculation;
it simply arises as a limiting case of our full result covering
all regions of fig.\ \ref{fig:overBH}b and the smooth transitions
between them.  The logarithm is a vacuum-like collinear logarithm, but the
limits of that collinearity (and so the argument of the logarithm)
are medium-dependent.
See the companion
paper \cite{softqed3} for a full discussion.%
\footnote{
  This type of contribution is sometimes referred to as
  ``direct'' pair production $e \to ee\bar e$.
  It was analyzed in certain limits
  by ref.\ \cite{BaierKatkov}, but we disagree with their analysis of
  the collinear logarithm.
}
That said, beyond the leading log, the limiting formulas for region
4 in Table \ref{tab:regions} represent a combination of
bremsstrahlung with and \textit{without} overlapping real pair
production, both contributing to $d\Gamma/dx_\gamma$.

\vspace{-0.1in}
\subsection*{Outlook}
\vspace{-0.1in}

The two black dots in figs.\ \ref{fig:overBH}--\ref{fig:LPM+overLPM} correspond
to two examples, with the most LPM suppression, of accurate measurements
of the LPM effect made at SLAC \cite{SLAC1,SLAC2} and CERN
\cite{CERNlpm1,CERNlpm2} by measuring the bremsstrahlung photon distribution
from electron beams on thin targets.
The black dots
are far from the $\LPMplus$ regions where the LPM effect is modified.
The CERN LPM experiment studied
electron energies up to 287 GeV, provided by an electron beam
\cite{H2beam}
created from
the 450 GeV proton beam of the SPS accelerator.
If, for example, someday a $\sim$50 TeV proton beam became available
(as in proposals for an FCC-hh accelerator \cite{FCC1,FCC2}),
and if the concept
of the CERN LPM experiment could be scaled up by a factor of 100 in electron
energy, that might open the possibility of exploring
fig.\ \ref{fig:LPM+overLPM} at the point marked by the small
open circle in fig.\ \ref{fig:LPM+overLPM},
where the $\LPMplus$ correction is significant.

However, for application to a future experiment of this type,
our calculations
will have to be refined to separate the contributions to
bremsstrahlung rates (i) with vs.\ (ii) without real pair production
(as done in ref.\ \cite{qedNfenergy} in the specific case
$E > \kgamma \gg \alpha E \gg \Elpma$).
Also, it will not be possible to make the
target simultaneously thick compared to the formation length yet thin
compared to the mean free time for pair production by the
bremsstrahlung photon; those two scales
are necessarily the same order of magnitude in the
$\LPMplus$ regions 4a and 4b.
That means that our calculations, which assume
an arbitrarily thick medium, would
have to be generalized to the finite-medium case.
These details should modify the quantitative analysis but not
the qualitative conclusion of $\LPMplus$ enhancement
relative to the original LPM effect.

Our qualitative expectation \cite{softqed3}
(not yet verified by explicit calculation)
is that,
for finite medium thickness $L \gtrsim 1/\Gamma_\pair$,
the boundary for $\LPMplus$ corrections to become
significant will remain parametrically the same as the region
$2|4$ boundary in table \ref{tab:boundaries} and fig.\ \ref{fig:LPM+overLPM},
but the parametric enhancement (\ref{eq:Sextra}) of the
$\LPMplus$/LPM ratio will be
restricted to never exceed $O(\Gamma_\pair L)$.
For $L \ll 1/\Gamma_\pair$, we expect that
the $\LPMplus$ effect will never be significant
since medium-induced
pair production by the bremsstrahlung photon then becomes unlikely.


\medskip
\begin{acknowledgments}

The work of Arnold and Bautista was supported,
in part, by the National Science Foundation under Grant No.~2412362.
Elgedawy was supported at different times by the
National Natural Science Foundation of China under Grant No.\ 12447145
and by the European Union (ERC, QGPthroughEECs, grant agreement
No.\ 101164102).
Views and opinions expressed are however those of the authors
only and do not necessarily reflect those of the European Union or the European
Research Council. Neither the European Union nor the granting authority can be
held responsible for them.
We are grateful to Spencer Klein and Ulrik Uggerh{\o}j for helpful discussions,
including Spencer's suggestion to consider FCC-hh and Ulrik's answers to
questions about the CERN LPM experiment.

\end{acknowledgments}

\textit{Data availability}---Mathematica code that generates numbers
used in the plots is openly available \cite{data}.


\appendix
\setcounter{equation}{0}\renewcommand\theequation{A\arabic{equation}}

\section{Appendix: full \boldmath$\protect\LPMplus$ rate formula}


See the companion paper \cite{softqed3} for derivation.
Our full formula for the $\LPMplus$ energy-loss rate
$d\Gamma/dx_\gamma = E\,d\Gamma/d\kgamma$ is
\begin {equation}
  \left[ \frac{d\Gamma}{dx_\gamma} \right]_\LPMplus
  =
  \left[ \frac{d\Gamma}{dx_\gamma} \right]_\LPM
  \left\{
     1 +
     \frac{ \delta \bigl[\frac{d\tilde\Gamma}{dx_\gamma}\bigr] }
          { \bigl[\frac{d\tilde\Gamma}{dx_\gamma}\bigr]_\LPM^{\rm soft} }
  \right\}
\end {equation}
where the overall factor of
$[d\Gamma/dx_\gamma]_\LPM$ is the full original
LPM rate (\ref{eq:LPMrate}) and the last term in braces is the
$\LPMplus$ correction to that rate.  We've expressed
the correction as a ratio of dimensionless rates
$d\tilde\Gamma \equiv (\Elpm/m^2) \, d\Gamma$.
The denominator of the correction term is the
soft-photon limit $x_\gamma \ll 1$ of (\ref{eq:LPMrate})
[the soft-photon limit being the only part
of fig.\ \ref{fig:LPM+overLPM} where the $\LPMplus$ correction is significant]:
\begin {equation}
  \left[ \frac{d\tilde\Gamma}{dx_\gamma} \right]_{\LPM}^{\rm soft} \!\!=
  \frac{\alpha}{\pi}
  \Re\left\{
      \sqrt{ \frac{2 i}{x_\gamma \vareps} } \,
        \biggl(
           1 - 2(b{+}b_\gamma)\,\PSI\bigl( 1;b{+}b_\gamma)
        \biggr)
  \right\} ,
\end {equation}
where $\vareps \equiv E/\Elpm$ is the initial electron energy in units of
$\Elpm$ and
\begin {equation}
   b \equiv \frac12 \sqrt{\frac{i x_\gamma}{2\vareps}}
   \qquad \mbox{and} \qquad
   b_\gamma \equiv \Bigl( \frac{m_\gamma}{x_\gamma m} \Bigr)^{\!2} b .
\end {equation}
The numerator of the $\LPMplus$ correction term is
\begin {equation}
  \delta \! \left[\frac{d\tilde\Gamma}{dx_\gamma}\right] =
  \left[\frac{d\tilde\Gamma}{dx_\gamma}\right]_\perp
  +
  \left[\frac{d\tilde\Gamma}{dx_\gamma}\right]_{\rm L} ,
\end {equation}
where the subscripts indicate contributions from processes
involving intermediate transversely ($\perp$)
or longitudinally (L) polarized photons, and
\begin {widetext}
\begin {multline}
  \left[\frac{d\tilde\Gamma}{dx_\gamma}\right]_\perp =
  \frac{\alpha^2}{\pi^2 x_\gamma} \,
  \Re \left\{
     \int_0^1 d\yfrak \>
     a
     \Bigl[ P_{\gamma\to e}(\yfrak)\,I_1(c,a) + a \, I_2(c,a) \Bigr]
  \right\}
\\
  - \frac{\alpha}{\pi} \,
  \Re\left\{
    \sqrt{ \frac{2i}{x_\gamma \vareps} } \,
    \Bigl[
      2(b'{+}b_\gamma) \bigl( \psi(1{+}b'{+}b_\gamma) + \gammaE \bigr)
      - 2b_\gamma \bigl( \psi(1{+}b_\gamma) + \gammaE \bigr)
    \Bigr]
  \right\} ,
\end {multline}
\begin {equation}
  \left[\frac{d\tilde\Gamma}{dx_\gamma}\right]_{\rm L} =
  - \frac{4\alpha^2}{\pi^2 x_\gamma}
    \Re\int_0^1 d\yfrak \> \yfrak(1{-}\yfrak) \, a \, I_3(a) .
\end {equation}
Above,
\begin {equation}
  a \equiv \sqrt{ \frac{ i }{ 2\yfrak(1{-}\yfrak)x_\gamma\vareps } } \,,
  \qquad
  b' \equiv \tilde{\cal G}_\pair \sqrt{\frac{\vareps}{2i x_\gamma}}
  \qquad
  c \equiv 2 x_\gamma \sqrt{\yfrak(1{-}\yfrak)} \,,
  \qquad
  P_{\gamma\to e}(\yfrak) = \yfrak^2 + (1{-}\yfrak)^2 ,
\end {equation}
and
\begin {equation}
  \tilde{\cal G}_\pair \equiv
  \frac{\alpha}{2\pi} \int_0^1 d\yfrak \>
  a
  \biggl\{
    P_{\gamma\to e}(\yfrak)
    \Bigl[
      1
      - a \, \PSI\bigl(1 \,; \tfrac{a}{2} \bigr)
    \Bigr]
    + a \, \PSI\bigl(\tfrac12 \,; \tfrac{a}{2}\bigr)
  \biggr\}
\end {equation}
is related to the LPM
pair production rate by $\tilde\Gamma_\pair = 2\Re(\tilde{\cal G}_\pair)$ .
The $I_n$ are the integrals
\begin {subequations}
\label {eq:In}
\begin{align}
  I_1(c,a) &\equiv
  \int_0^\infty d\tau\>
  \Bigl( \frac{1}{\sinh^2\tau} - \frac{1}{\tau^2} \Bigr)
  \ln(c\tau) \, e^{-a\tau}
\nonumber\\ & \qquad =
  \left[
     a\, \PSI\bigl(1;\tfrac{a}{2}\bigr)
     - 1
  \right]
  \left[ \ln\bigl(\tfrac{c}{2}\bigr) - \gammaE \right]
  + a
    \left[
       \gamma_1\bigl(\tfrac{a}{2}\bigr)
       + \tfrac12 \ln^2\bigl(\tfrac{a}{2}\bigr)
       - \ln\bigl(\tfrac{a}{2}\bigr) + 1
    \right]
  + 2\lnGamma\bigl(\tfrac{a}{2}\bigr) - \ln(2\pi) ,
\label {eq:I1}
\\
  I_2(c,a) &\equiv
  \int_0^\infty d\tau\>
  \Bigl( \frac{1}{\sinh\tau} - \frac{1}{\tau} \Bigr)
  \ln(c\tau) \, e^{-a\tau}
  =
  -
     \PSI\bigl(\tfrac12;\tfrac{a}{2}\bigr)
  \left[ \ln\bigl(\tfrac{c}{2}\bigr) - \gammaE \right]
  - \gamma_1\bigl(\tfrac12{+}\tfrac{a}{2}\bigr)
  - \tfrac12 \ln^2\bigl(\tfrac{a}{2}\bigr) ,
\\
  I_3(a)\phantom{c,} &\equiv
  \int_0^\infty \frac{d\tau}{\tau} \>
  \left( \frac{1}{\sinh\tau} - \frac{1}{\tau} \right) e^{-a\tau}
  =
  2 \Bigl[
      \lnGamma\bigl(\tfrac12{+}\tfrac{a}{2}\bigr)
      - \tfrac{a}{2} \ln\bigl(\tfrac{a}{2}\bigr)
      + \tfrac{a}{2}
      - \tfrac12 \ln(2\pi)
    \Bigr]
\end {align}
\end {subequations}
with again $\PSI(r;z) \equiv \psi(r{+}z) - \ln z$.
\end {widetext}

The special function $\gamma_1$ in (\ref{eq:In})
is the $n{=}1$ case of the
generalized Stieltjes constants defined by the Laurent series expansion
of the Hurwitz zeta function about its $s{=}1$ pole:
\begin {equation}
  \zeta(s,q) \equiv \sum_{k=0}^\infty \frac{1}{(k+q)^s}
  = \frac{1}{s-1}
    + \sum_{n=0}^\infty \, \frac{(-)^n}{n!} \, \gamma_n(q) \, (s{-}1)^n .
\end {equation}
The log-Gamma function $\lnGamma(z)$ in (\ref{eq:In}) is
defined with the convention that all cuts run along the negative
real $z$ axis.  In contrast, with the standard choice of cut for $\ln(w)$,
taking $\ln\bigl(\Gamma(z)\bigr)$ would give incorrect (and discontinuous)
results in our application;
so one must use an appropriate, direct implementation of
$\lnGamma(z)$ in numerical calculations.
There exist numerical implementations of both $\gamma_n(q)$ and
$\lnGamma(z)$.%
\footnote{
  e.g.\ StieltjesGamma[$n$,$q$] and LogGamma[$z$] in Mathematica.
}



\begin{thebibliography}{}

\bibitem{BH}
  H.~Bethe and W.~Heitler,
  ``On the Stopping of fast particles and on the creation of
    positive electrons,''
  Proc. Roy. Soc. Lond. A \textbf{146}, 83-112 (1934)
  doi:10.1098/rspa.1934.0140

\bibitem{LP1}
  L.~D.~Landau and I.~Pomeranchuk,
  ``Limits of applicability of the theory of bremsstrahlung electrons and
  pair production at high-energies,''
  Dokl.\ Akad.\ Nauk Ser.\ Fiz.\  {\bf 92} (1953) 535

\bibitem{LP2}
  L.~D.~Landau and I.~Pomeranchuk,
  ``Electron cascade process at very high energies,''
  Dokl. Akad. Nauk Ser. Fiz.  {\bf 92} (1953) 735

\bibitem{LPenglish}
  L. Landau,
  {\sl The Collected Papers of L.D. Landau}\/
  (Pergamon Press, New York, 1965)

\bibitem{Migdal}
  A.~B.~Migdal,
  ``Bremsstrahlung and pair production in condensed media at high energies,''
   Phys. Rev.  {\bf 103}, 1811 (1956)

\bibitem{SLAC1}
  P.~L.~Anthony, R.~Becker-Szendy, P.~E.~Bosted, M.~Cavalli-Sforza,
  L.~P.~Keller, L.~A.~Kelley, S.~R.~Klein, G.~Niemi, M.~L.~Perl
  and L.~S.~Rochester,
  \textit{et al.}
  ``An Accurate measurement of the Landau-Pomeranchuk-Migdal effect,''
  Phys. Rev. Lett. \textbf{75}, 1949-1952 (1995)
  doi:10.1103/PhysRevLett.75.1949

\bibitem{SLAC2}
  P.~L.~Anthony \textit{et al.} [SLAC-E-146],
  ``Bremsstrahlung suppression due to the LPM and dielectric effects
    in a variety of materials,''
  Phys. Rev. D \textbf{56}, 1373-1390 (1997)
  doi:10.1103/PhysRevD.56.1373

\bibitem{CERNlpm1}
  H.~D.~Hansen, U.~I.~Uggerh{\o}j, C.~Biino, S.~Ballestrero, A.~Mangiarotti,
  P.~Sona, T.~J.~Ketel and Z.~Z.~Vilakazi,
  ``Is the electron radiation length constant at high energies?,''
  Phys. Rev. Lett. \textbf{91}, 014801 (2003)
  doi:10.1103/PhysRevLett.91.014801

\bibitem{CERNlpm2}
  H.~D.~Hansen, U.~I.~Uggerh{\o}j, C.~Biino, S.~Ballestrero, A.~Mangiarotti,
  P.~Sona, T.~J.~Ketel and Z.~Z.~Vilakazi,
  ``Landau-Pomeranchuk-Migdal effect for multihundred GeV electrons,''
  Phys. Rev. D \textbf{69}, 032001 (2004)
  doi:10.1103/PhysRevD.69.032001

\bibitem{SpencerReview}
  S.~Klein,
  ``Suppression of bremsstrahlung and pair production due to
    environmental factors,''
  Rev. Mod. Phys. \textbf{71}, 1501-1538 (1999)
  doi:10.1103/RevModPhys.71.1501
  [arXiv:hep-ph/9802442 [hep-ph]]

\bibitem{Galitsky}
  V.~M.~Galitsky and I.~I.~Gurevich,
  ``Coherence effects in ultra-relativistic electron bremsstrahlung,''
  Nuovo Cimento \textbf{32}, 396 (1964)

\bibitem{softqed3}
  P.~Arnold, J.~Bautista, O.~Elgedawy and S.~Iqbal,
  ``Calculating extremely high energy bremsstrahlung in matter,''
  [arXiv:2604.18685 [hep-ph]]

\bibitem{softqed1}
 P.~Arnold, J.~Bautista, O.~Elgedawy and S.~Iqbal,
  ``Revisiting extremely high energy QED bremsstrahlung in matter:
    large modifications to the LPM effect,''
  JHEP \textbf{03}, 015 (2026)
  doi:10.1007/JHEP03(2026)015
  [arXiv:2508.21120 [hep-ph]]

\bibitem{RPP2024}
  S.~Navas \textit{et al.}\ [Particle Data Group],
  ``Review of particle physics,''
  Phys. Rev. D \textbf{110}, no.3, 030001 (2024)
  doi:10.1103/PhysRevD.110.030001

\bibitem{LPMphotoKlein}
  L.~Gerhardt and S.~R.~Klein,
  ``Electron and Photon Interactions in the Regime of Strong LPM Suppression,''
  Phys. Rev. D \textbf{82}, 074017 (2010)
  doi:10.1103/PhysRevD.82.074017
  [arXiv:1007.0039 [hep-ph]]

\bibitem{BaierKatkov}
  V.~N.~Baier and V.~M.~Katkov,
  ``Electroproduction of electron-positron pair in a medium,''
  JETP Lett. \textbf{88}, 80-84 (2008)
  doi:10.1134/S0021364008140026
  [arXiv:0805.0456 [hep-ph]]

\bibitem{H2beam}
  N.~Charitonidis and I.~Efthymiopoulos,
  ``Low energy tertiary beam line design for the
    CERN neutrino platform project,''
  Phys. Rev. Accel. Beams \textbf{20}, no.11, 111001 (2017)
  doi:10.1103/PhysRevAccelBeams.20.111001

\bibitem{FCC1}
  M.~Benedikt \textit{et al.} [FCC],
  ``Future Circular Collider Feasibility Study Report:
    Volume 1, Physics, Experiments, Detectors,''
  Eur. Phys. J. C \textbf{85}, no.12, 1468 (2025)
  doi:10.1140/epjc/s10052-025-15077-x
  [arXiv:2505.00272 [hep-ex]]

\bibitem{FCC2}
  M.~Benedikt \textit{et al.} [FCC],
  ``Future Circular Collider Feasibility Study Report:
    Volume 2, Accelerators, Technical Infrastructure and Safety,''
  Eur. Phys. J. ST \textbf{234}, no.19, 5713-6197 (2025)
  doi:10.1140/epjs/s11734-025-01967-4
  [arXiv:2505.00274 [physics.acc-ph]]

\bibitem{qedNfenergy}
  P.~Arnold, O.~Elgedawy and S.~Iqbal,
  ``Strong vs.\ weakly coupled in-medium showers: energy stopping in
    large-$\Nf$ QED,''
  arXiv:2404.19008 [hep-ph]

\bibitem{data}
  https://doi.org/10.5281/zenodo.22283988

\end{thebibliography}
\end{document}